\documentclass[%
 reprint,
 amsmath,amssymb,
 aps,
nofootinbib]{revtex4-1}

\usepackage{graphicx}% Include figure files
\usepackage{dcolumn}% Align table columns on decimal point
\usepackage{bm}% bold math
\usepackage{hyperref}
\usepackage{float}
\usepackage{xcolor}

\begin{document}
\setlength{\abovedisplayskip}{5pt}
\setlength{\belowdisplayskip}{5pt}
\setlength{\abovedisplayshortskip}{5pt}
\setlength{\belowdisplayshortskip}{5pt}

\preprint{}

\title{Plausible Indication of Gamma-Ray Absorption by Dark Matter in NGC 1068}

\author{Gonzalo Herrera}
\affiliation{Center for Neutrino Physics, Department of Physics,\\ Virginia Tech, Blacksburg, VA 24061, USA}
\email{gonzaloherrera@vt.edu}

\begin{abstract}
NGC 1068 is the brightest extragalactic source in high-energy neutrinos as seen by IceCube, yet the accompanying gamma-ray flux is orders of magnitude weaker. It has been argued that this indicates that the bulk of neutrinos and gamma rays are emitted in the innermost vicinity of the central supermassive black hole, which is transparent to neutrinos, but opaque to gamma rays. Even in such extreme scenarios for the acceleration of cosmic rays, astrophysical models typically overestimate the low-energy gamma-ray flux and/or require some fine-tuning in the physical parameters. Here we suggest instead that the dark matter surrounding the supermassive black hole may absorb the gamma rays, inducing the observed deficit. We show that for a dark matter-photon scattering cross section in the range $\sigma_{\rm DM-\gamma}/m_{\rm DM} \simeq 10^{-28}-10^{-30}$ cm$^2$/GeV, Fermi-LAT measurements can be well reconciled with IceCube data. We also present some simple particle physics examples that achieve the correct spectral energy dependence while respecting complementary constraints.
\end{abstract}

\maketitle

%\tableofcontents

\section{Introduction}
The IceCube collaboration observed $79_{-20}^{+22}$ high-energy neutrinos in the direction of NGC 1068 at $4.2\sigma$  \cite{IceCube:2022der}. On the other hand, the MAGIC and Fermi-LAT collaborations did not observe the accompanying very high-energy gamma-ray flux \cite{MAGIC:2019fvw}, and the gamma-ray flux observed by Fermi-LAT is orders of magnitude weaker than the observed high-energy neutrino flux \cite{2012ApJ...755..164A}. It has been suggested that the deficit of gamma rays indicates that the emission arises from cosmic rays accelerated in the innermost vicinity of the black hole, where the attenuation of the gamma-ray flux due to scatterings with lower-energy ambient photons can be significant \cite{Inoue:2022yak,Murase:2022dog, Blanco:2023dfp, Das:2024vug, Eichmann_2022, Fiorillo:2023dts, Fiorillo:2024akm, Yasuda:2024fvc}. However, even in such arguably extreme scenarios for the acceleration of cosmic rays, standard single-zone emission lepto-hadronic models typically overestimate low-energy gamma rays compared to the observations by Fermi-LAT \cite{2020ApJS..247...33A}. In this work we suggest that dark matter surrounding the supermassive black hole of NGC 1068 may instead be responsible for the observed deficit of gamma rays. We begin by quantifying the absorption coefficient needed to reconcile the expected gamma-ray fluxes from $pp$ and $p\gamma$ models with Fermi-LAT measurements. Then, we compute the column density of dark matter in NGC 1068, and from that we infer a model-independent dark matter-photon scattering cross section required to induce the additional absorption. Finally, we present some simple particle physics models where the necessary photon absorption can be accommodated with the correct spectral dependence with energy. We further discuss that some regions of the preferred parameter space fulfill the constraints on the dark matter-photon interaction cross-section from cosmology, colliders, and from the other significant Active Galactic Nuclei (AGN) observed in high-energy neutrinos, TXS 0506+056 \cite{IceCube:2018dnn}.
	\begin{figure*}[t!]
		\centering
		\includegraphics[width=0.475\textwidth]{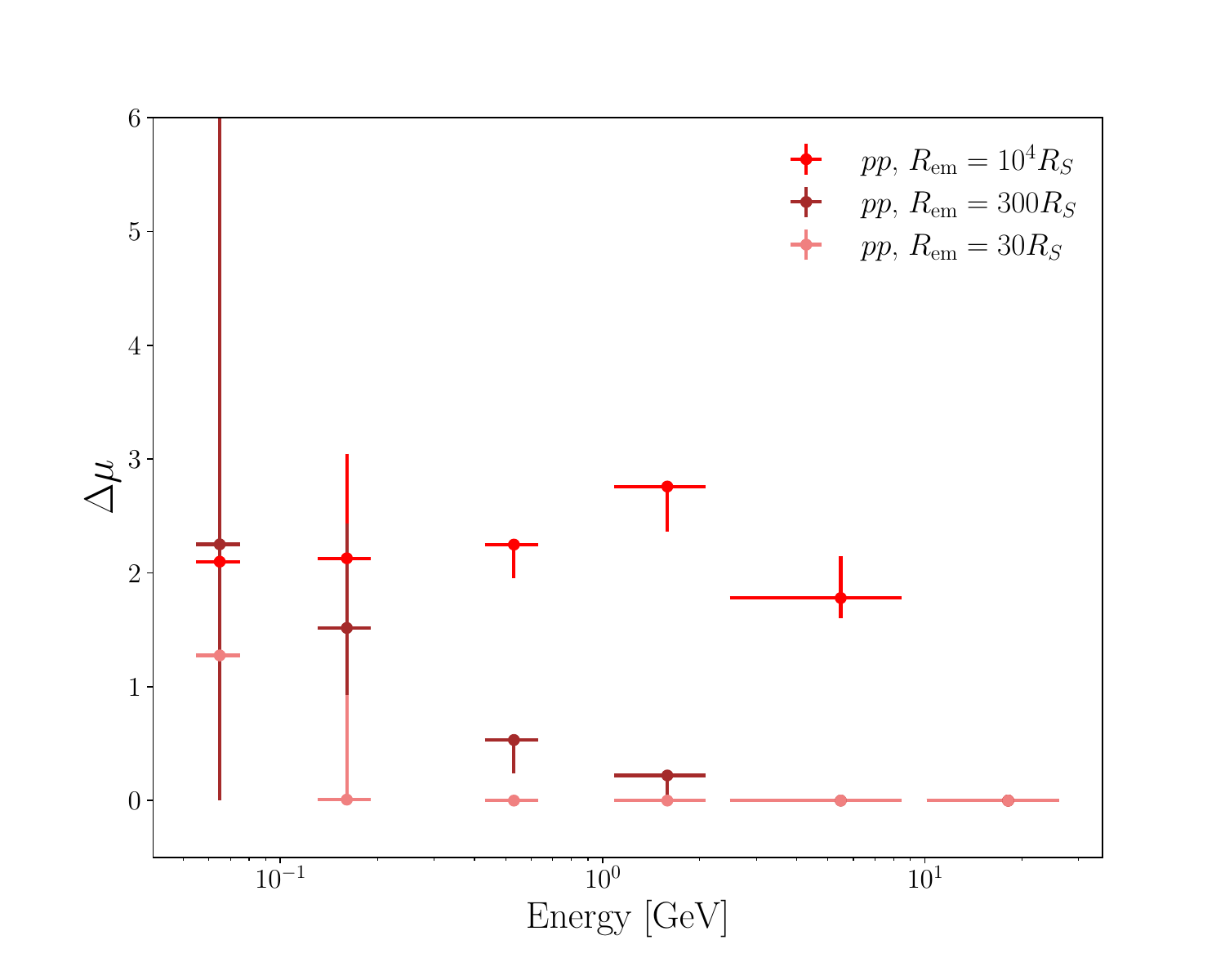}
  		\includegraphics[width=0.475\textwidth]{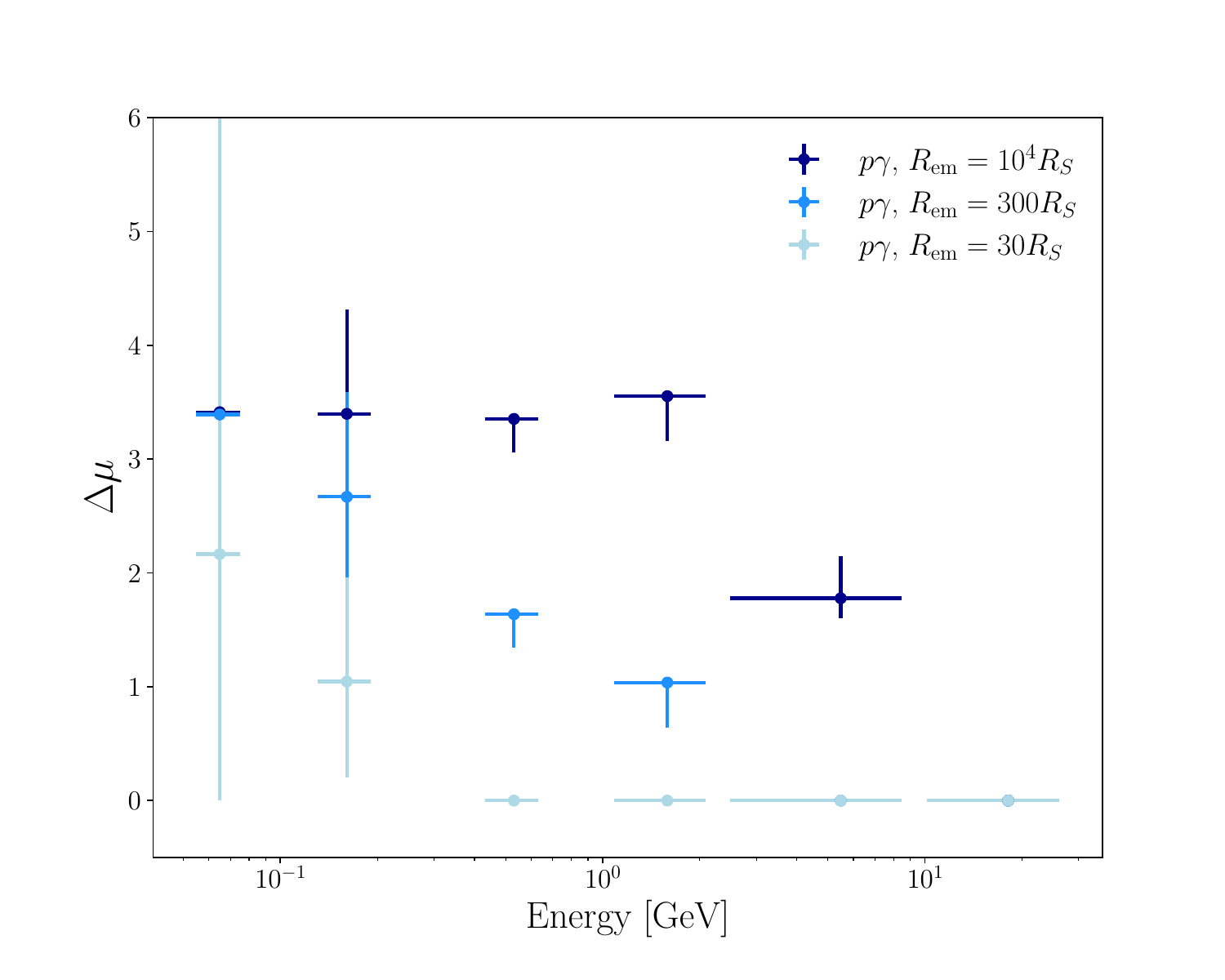}
		\centering
  		\caption{Absorption coefficient needed to attenuate the expected gamma-ray flux from NGC 1068 \cite{Inoue_2020, Murase:2022dog} to the observed values by Fermi-LAT (including error bars) \cite{2012ApJ...755..164A, 2020ApJS..247...33A}, for both $pp$ (left panel) and $p\gamma$ (right panel) scenarios and different locations of the emitting region of neutrinos and gamma rays. Absorption coefficients larger than these values would be compatible with Fermi-LAT data as well, if including an additional component of gamma rays arising from the starburst activity with high supernova rates \cite{Eichmann_2022}.}
	\label{fig:attenuation-coeff}
    \end{figure*}

\section{Gamma-ray and neutrino fluxes from NGC 1068}\label{sec:fluxes}
In lepto-hadronic models, high-energy neutrinos and gamma rays are simultaneously produced by meson decays in $p\gamma$ and $pp$ interactions \cite{Berezinsky:1975zz,Stecker:1991vm, Murase_2016}, with comparable fluxes in neutrinos and gamma rays. The gamma-ray flux from NGC 1068 has been measured by Fermi-LAT in the $0.1-100$ GeV energy range \cite{2012ApJ...755..164A,2020ApJS..247...33A}, while MAGIC has placed upper limits at sub-TeV energies \cite{MAGIC:2019fvw}. On the other hand, the IceCube collaboration recently reported a $4.2 \sigma$ measurement of neutrinos around 1 TeV, significantly above the upper limits placed by MAGIC and the gamma-ray flux measured by Fermi-LAT \cite{IceCube:2022der}. A potential solution to the discrepancy between traditional lepto-hadronic single-zone models and the telescope data relies on high-energy gamma rays and neutrinos being produced near the central supermassive black hole of NGC 1068, at the coronal region $\sim 30 R_S - 300 R_S$ \cite{Inoue_2020, Inoue:2022yak, Murase:2022dog, Blanco:2023dfp, Eichmann_2022}, with $R_S$ denoting the Schwarzschild radius. However, these models are unable to reproduce the \textit{full} spectrum measured by Fermi-LAT for $pp$ or $p\gamma$ production, failing particularly at the lowest and highest energy bins measured by the collaboration. For the lowest energy bins, these models typically overestimate photons compared to Fermi-LAT data. At the highest energy bins, they can't produce a sufficiently large gamma-ray flux, and starburst activity with high supernova rates has been invoked as a potential source of such energetic gamma rays \cite{Eichmann_2022, Romeo:2016hms, Ajello:2023hkh}. However, neutrinos and gamma rays could also be produced farther from the central black hole, at the typical location of the Broad Line Region (BLR) $\sim 10^{4} R_S$, or beyond. In fact, this is the expectation in traditional lepto-hadronic single-zone emission models proposed to explain the only extragalactic high-energy neutrino source known besides NGC 1068, TXS 0506+056 \cite{Padovani_2019}. In this case, the expected gamma-ray flux from NGC 1068 might still be compatible at the highest energies measured by Fermi-LAT, but current models still overestimate by orders of magnitude the gamma-ray flux in the energy range $\sim 0.07-5$ GeV. The lack of sensitivity to photon emission from these sources in the MeV energy range makes it more difficult to constrain the emitting region of neutrinos and gamma rays from NGC 1068 within lepto-hadronic models.

The gamma-ray flux at GeV-TeV energies can be attenuated by the accretion disk, the corona, the BLR and the dust torus. The optical depth $\tau$ to electron-positron pair production ($\gamma\gamma \rightarrow e^{-}e^{+}$) at TeV energies can reach high values $\rm log(\tau) \sim 1-5$, such that these photons can hardly escape the source even when they are produced at the BLR $\sim 10^{4}R_S$. Further, gamma rays could also be attenuated due to interactions with the Extragalactic Background Light, although the optical depth in this case only becomes relevant at TeV energies \cite{Cooray:2016jrk}. 
Additionally, gamma rays could also be attenuated at different energies due to Beyond the Standard Model (BSM) interactions. Here we will focus on the putative interactions between photons and the ambient dark matter particles present in the vicinity of NGC 1068. For the moment, we remain agnostic about the source of attenuation, and discuss the absorption coefficient needed to reconcile NGC 1068 gamma-ray and neutrino production models with the Fermi-LAT data in every energy bin. In particular, we wind the value of $\Delta\mu$ such that:
\begin{align}
\frac{\Phi^{\rm obs}(E_{\gamma})}{\Phi^{\rm em}(E_{\gamma})}= e^{-(\mu_{\gamma \gamma}+\Delta\mu(E_\gamma))}
\end{align}
holds, where $\Phi^{\rm em}$ is the emitted gamma-ray spectrum in lepto-hadronic models, $\Phi^{\rm obs}$ is the observed spectrum by Fermi-LAT, and $\mu_{\gamma\gamma}$ is the optical depth of electron-positron production. The expected gamma-ray flux depends on the emitting region $R_{\rm em}$, as well as on the magnetic field of the AGN, parametrized by $\epsilon_{B}=B^2 f_{\Omega}R_{\rm em}^2c/2L_{\rm bol}$, where $B$ is the magnetic field strength, $R_{\rm em}$ is the emitting region of gamma rays, $f_{\Omega}=\Delta \Omega/4\pi$ is a geometrical factor and $L_{\rm bol}$ is the bolometric luminosity of the AGN \cite{Murase:2022dog}. For NGC 1068, $L^{\rm NGC}_{\mathrm{bol}}\simeq10^{45} \mathrm{erg} \mathrm{~s}^{-1}$ \cite{Woo:2002un}. We will consider production via $pp$ and $p\gamma$ processes, and assume the $\epsilon_{B}=0.01$ for the $pp$ scenario, and $\epsilon_{B}=1$ for the $p\gamma$ case, according to \cite{Murase:2022dog}. In principle, both $pp$ and $p\gamma$ processes could be operational simultaneously, yielding comparable neutrino and gamma-ray fluxes. However, we will treat both scenarios independently as the dominant mechanism responsible for neutrino and gamma-ray emission, as described in \cite{Murase:2022dog}. This corresponds to the case where the Inverse Compton cascade processes dominates for $pp$ production, while the contribution of Bethe-Heitler pair production becomes relevant for $p\gamma$ processes. The values chosen for the magnetic field strength are model-dependent, however, larger values for $pp$ as $\epsilon_{B}=1$ do not allow to reconcile the expected fluxes with Fermi-LAT data at the highest energy bins, only mildly improving the agreement in the lower part of the spectrum for emission beyond $R_{\rm em} \gtrsim 10^{4}R_S$.
	 \begin{figure*}[t!]
		\centering
		\includegraphics[width=0.8\textwidth]{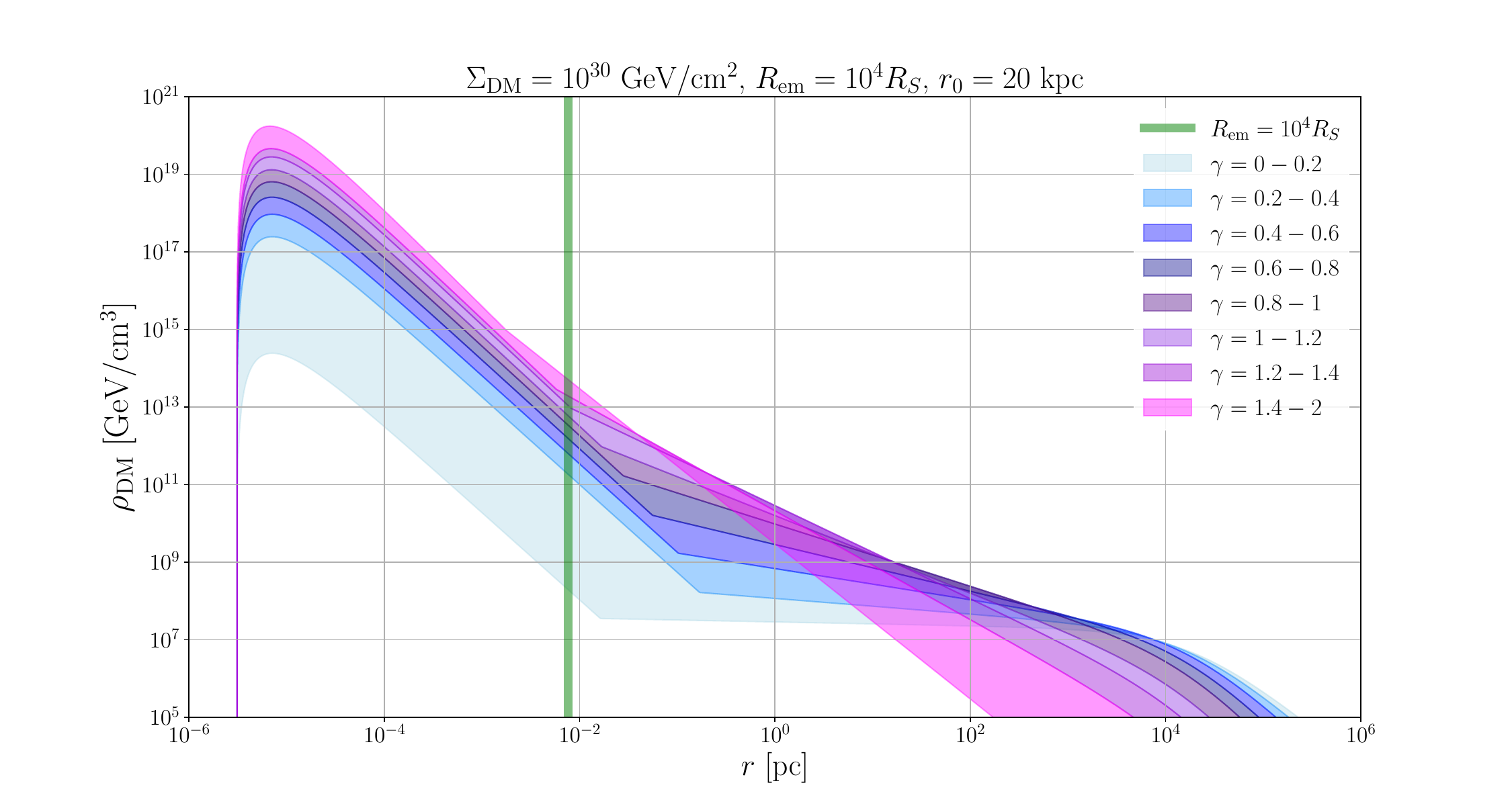}
		\centering
		\caption{Family of dark matter distributions around the supermassive black hole of NGC 1068 yielding a column density of $\Sigma_{\rm DM}=10^{30}$ GeV/cm$^2$, assuming emission from $R_{\rm em}=10^4 R_S$, and for different values of the initial NFW-like profile index $\gamma$.} 
		\label{fig:DMspike}
	\end{figure*}
    
In Figure \ref{fig:attenuation-coeff}, we show the lower limit on the absorption coefficient needed in every energy bin measured by Fermi-LAT, in both $pp$ and $p\gamma$ production mechanisms, and for different choices of the emitting region of neutrinos and gamma rays \cite{Murase:2022dog}. We perform the calculation for the central value of Fermi-LAT as well as for the upper and lower error bars reported by the collaboration. For models with emission from the coronal region $\sim 30R_S-300R_S$, the expected fluxes at the highest energy bins measured by Fermi-LAT are well below the observed ones. Here, no absorption is needed, on the other hand, an additional source of high energy photons is required to fit the data. As previously discussed, one possibility is to invoke an enhanced starburst activity farther from the supermassive black hole and with high supernova rates ($\sim 0.5$ yr$^{-1}$),  \cite{Eichmann_2022}.

\section{Dark matter distribution in the vicinity of NGC 1068}
Adiabatically-growing black holes are expected to form a dense spike of dark matter around them \cite{1972GReGr...3...63P,Quinlan:1994ed,Gondolo_1999, Ullio:2001fb}. An initial profile of the form $\rho (r) = \rho_0 (r/r_0)^{-\gamma}$ evolves into
\begin{align}
	\rho_{\rm sp}(r) = \rho_{R} \, g_{\gamma}(r)\, \Big(\frac{R_{sp}}{r}\Big)^{\gamma_{\rm sp}}\;,
\end{align}
where $R_{\rm sp}=\alpha_{\gamma}r_0(M_{\rm BH}/(\rho_{0}r_{0}^{3})^{\frac{1}{3-\gamma}}$ is the size of the spike,  and $\gamma_{\rm sp}=\frac{9-2\gamma}{4-\gamma}$ parametrizes the cuspiness of the spike. Further, $g_{\gamma}(r)$ is a function which  can be approximated for $0<\gamma <2 $ by  $g_{\gamma}(r) \simeq (1-\frac{4R_{S}}{r})^3$, with $R_S$ the Schwarzschild radius, while $\rho_{\rm R}$ is a normalization factor, chosen to match the density profile outside of the spike, $\rho_R=\rho_{0}\, (R_{sp}/r_0)^{-\gamma}$. This density profile is defined only for $r\gtrsim 4 R_S$; for smaller radial coordinates, the density profile vanishes.

In the following, we will assume that far away from the black hole, the dark matter distribution follows the standard NFW profile~\cite{Navarro:1995iw}. We will consider values of $\gamma$ ranging from 0 to 2, with values ranging from $\alpha_{\gamma}= 0.00733-0.0177$. The mass of the black hole at the center of NGC 1068 was estimated from the rotational motion of a water maser disk to be $M_{\rm BH}= 0.8-1.7 \times 10^{7} M_{\odot}$ \cite{Woo:2002un, Panessa:2006sg}, although values as large as $M_{\rm BH}\approx 1 \times 10^{8} M_{\odot}$ have been obtained from the neutral FeK$\alpha$ line \cite{Minezaki_2015}. In the following, we will assume the mass of the black hole of NGC 1068 to be $M_{\rm BH}= 0.8 \times 10^{7} M_{\odot}$. We adopt the lower end of the water maser-based black hole mass measurements because maser kinematics directly trace Keplerian rotation on sub-parsec scales, providing a robust dynamical estimate. In contrast, Fe K$\alpha$-based masses rely on emission originating from much larger, reflection-dominated regions ($\sim 10-100$ pc) and are therefore more model-dependent. As we will demonstrate later on, the mass of the central black hole will affect the photon absorption coefficient from dark matter with a scaling of $ \propto M_{\rm BH}^{3/4}$, so the uncertainty in the water maser black hole mass determinations from NGC 1068 affects the absorption coefficient by at most a factor of $\sim 1.8$. The Schwarzschild radius is  $R_S\approx 7.7 \times 10^{-7}$ pc. We have taken $r_0$=20 kpc, in analogy to the Milky Way. At this point of the discussion, we leave the normalization $\rho_0$ as a free parameter. 

This profile is only valid when the dark matter particles  do not annihilate ({\it e.g} as in scenarios of asymmetric dark matter), or do so very slowly. Otherwise, the maximal dark matter density in the inner regions of the spike is saturated to $\rho_{\text {sat }} = m_{\rm DM} /(\langle\sigma v \rangle t_{\mathrm{BH}})$, where $\langle \sigma v \rangle$ is the velocity averaged dark matter annihilation cross section, and $t_{\rm BH}$ is the time elapsed since the black hole formation, for which we take the value $t_{\rm BH}=10^{10}$ yr. Further, the dark matter profile of the spike extends to a maximal radius $R_{\rm sp}$, beyond which the dark matter distribution follows the pre-existing NFW profile. In full generality, the dark matter profile in the spike reads \cite{Gondolo_1999,Lacroix_2015})
\begin{align}\rho(r)= \begin{cases} 
		0 & r\leq 4R_S \\
		\frac{\rho_{\rm sp}(r)\rho_{\rm sat}}{\rho_{\rm sp}(r)+\rho_{\rm sat}} & 4R_S\leq r\leq R_{sp} \\
		\rho_{0}\Big(\frac{r}{r_0}\Big)^{-\gamma} \Big(1+\frac{r}{r_0}\Big)^{-2} & r\geq R_{sp} .
	\end{cases}
	\label{eq:spike_profile}
\end{align}

A set of plausible dark matter profiles in NGC 1068 is shown in the left panel of Figure \ref{fig:DMspike} for various values of $\gamma$, and values of the self-annihilation cross section that leave the dark matter spike intact, of $\langle \sigma v\rangle \lesssim 10^{-37}$cm$^2$. Here the parameter $\rho_0$ is fixed for each case such that the column density from $R_{\rm em}=10^{4}R_S$ yields $\Sigma_{\rm DM}=10^{30}$GeV/cm$^2$. This reproduces well the observed halo mass of the host Galaxy of NGC 1068 ($M_{\rm DM} \simeq 10^{12}M_{\odot}$). As apparent from the plot, the dark matter density is extremely high at the location where neutrinos and gamma rays are expected to be produced, corresponding to the left-hand side of the vertical green colored line. Thus, interactions with dark matter particles may occur with sufficient frequency to produce a sizable attenuation of the flux.
\begin{figure*}[t!]
\centering

    \includegraphics[width=0.45\textwidth]{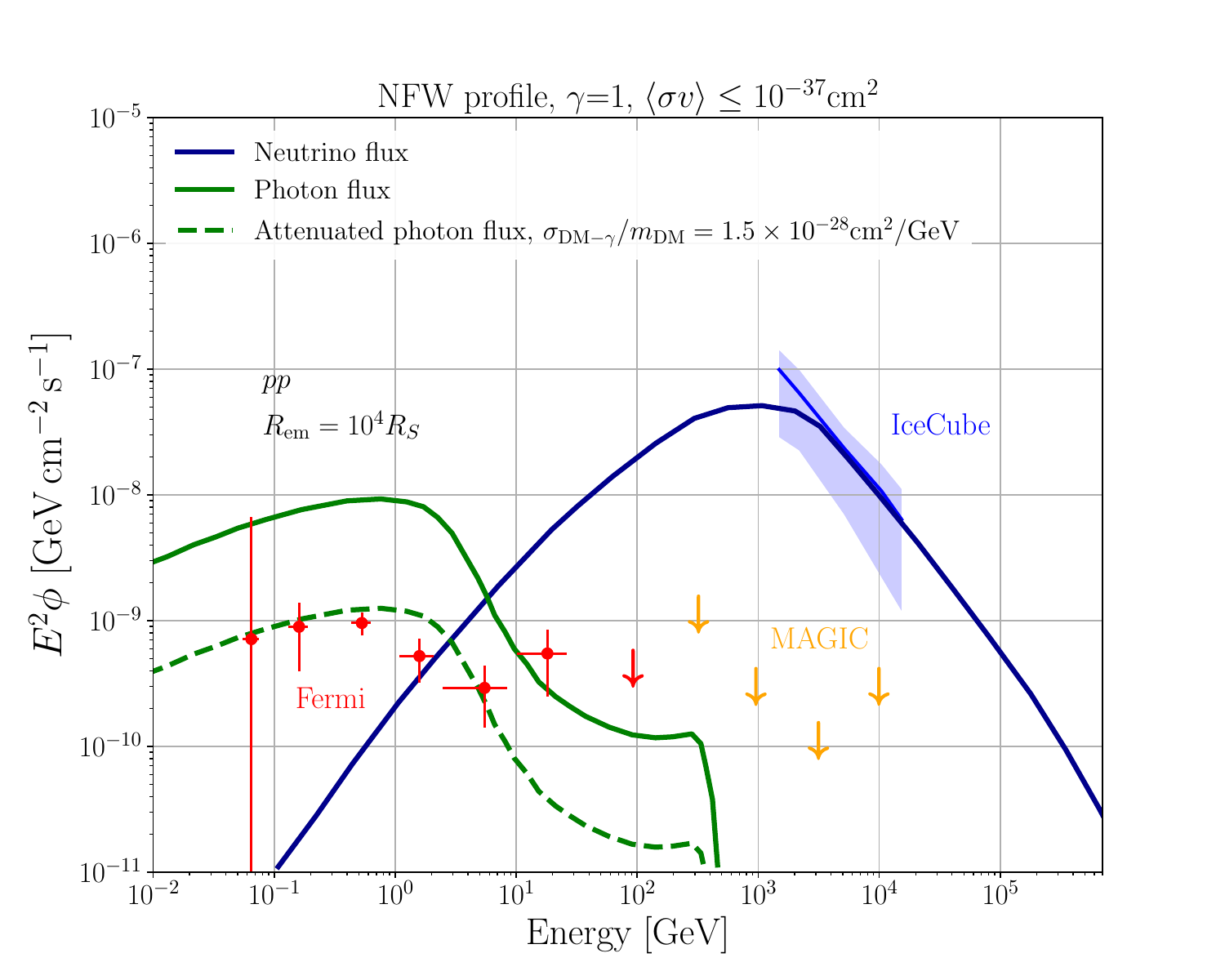}
     \includegraphics[width=0.45\textwidth]{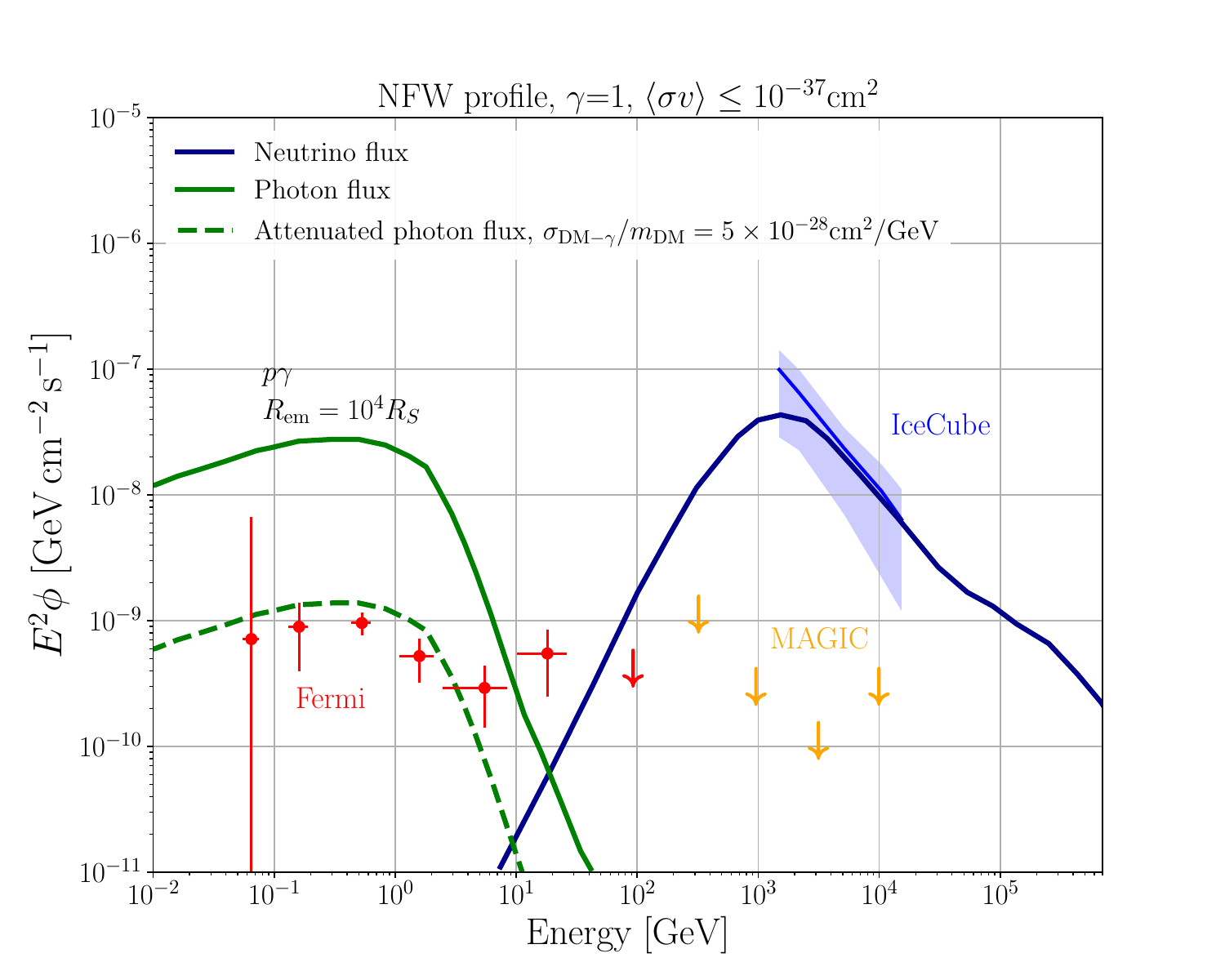}

		\caption{High-energy neutrino and gamma-ray spectral distribution from NGC 1068, including IceCube \cite{IceCube:2022der}, Fermi-LAT \cite{2012ApJ...755..164A, 2020ApJS..247...33A}, and MAGIC \cite{MAGIC:2019fvw} data. The solid blue lines reflect the predicted high-energy neutrino flux via $pp$ (left panel) and $p\gamma$ (right panel) interactions at $R_{\rm em} \simeq 10^{4}R_S$ \cite{Murase:2022dog}. The solid green lines show the corresponding gamma-ray flux, which overshoots Fermi-LAT measurements at most energy bins. The dashed dotted lines show the attenuated photon flux due to absorption by dark matter particles, for different values of the cross section, and the parameters of the dark matter spike as normalized in Eq. \ref{eq:absorption_coeff} with $\gamma=1$.}
		\label{fig:attenuation_constant}
\end{figure*}
We turn now into computing the photon absorption coefficient induced by dark matter. This is given by
\begin{align}
\Delta \mu\big|_{\rm DM}=\frac{\sigma_{\rm DM-\gamma}(E_{\gamma})\Sigma_{\rm DM}}{m_{\rm DM}}
\end{align}
where $\sigma_{\rm DM-\gamma}(E_{\gamma})$ is the dark matter-photon scattering cross section, which in general depends on the photon energy, and $\Sigma_{\rm DM}$ is the number of dark matter particles along the path of photons
\begin{align}
\Sigma_{\rm DM}=\int_{\rm path} dr\rho(r)
\label{eq:Sigma}
\end{align}
In this paper we focus on the impact on the attenuation of the passage through the dark matter in NGC 1068, with density profile given in Eq.~(\ref{eq:spike_profile}), and that it is orders of magnitude stronger than the contribution to $\Sigma_{\rm DM}$ from the dark matter in the intergalactic medium and in the Milky Way.
We will then approximate
\begin{align}
\Sigma_{\rm DM}\simeq 
\Sigma_{\rm DM}\Big|_{\rm sp} +
\Sigma_{\rm DM}\Big|_{\rm host}\simeq \int_{R_{\rm em}}^{R_{\rm sp}} dr\rho(r)+\int_{R_{\rm sp}}^{10r_0} dr\rho(r)\;.
\end{align}
where the contribution from the spike will generically dominate over the contribution from the galactic halo \cite{Ferrer:2022kei}.

\begin{figure*}[t!]
		\centering
        \includegraphics[width=0.45\textwidth]{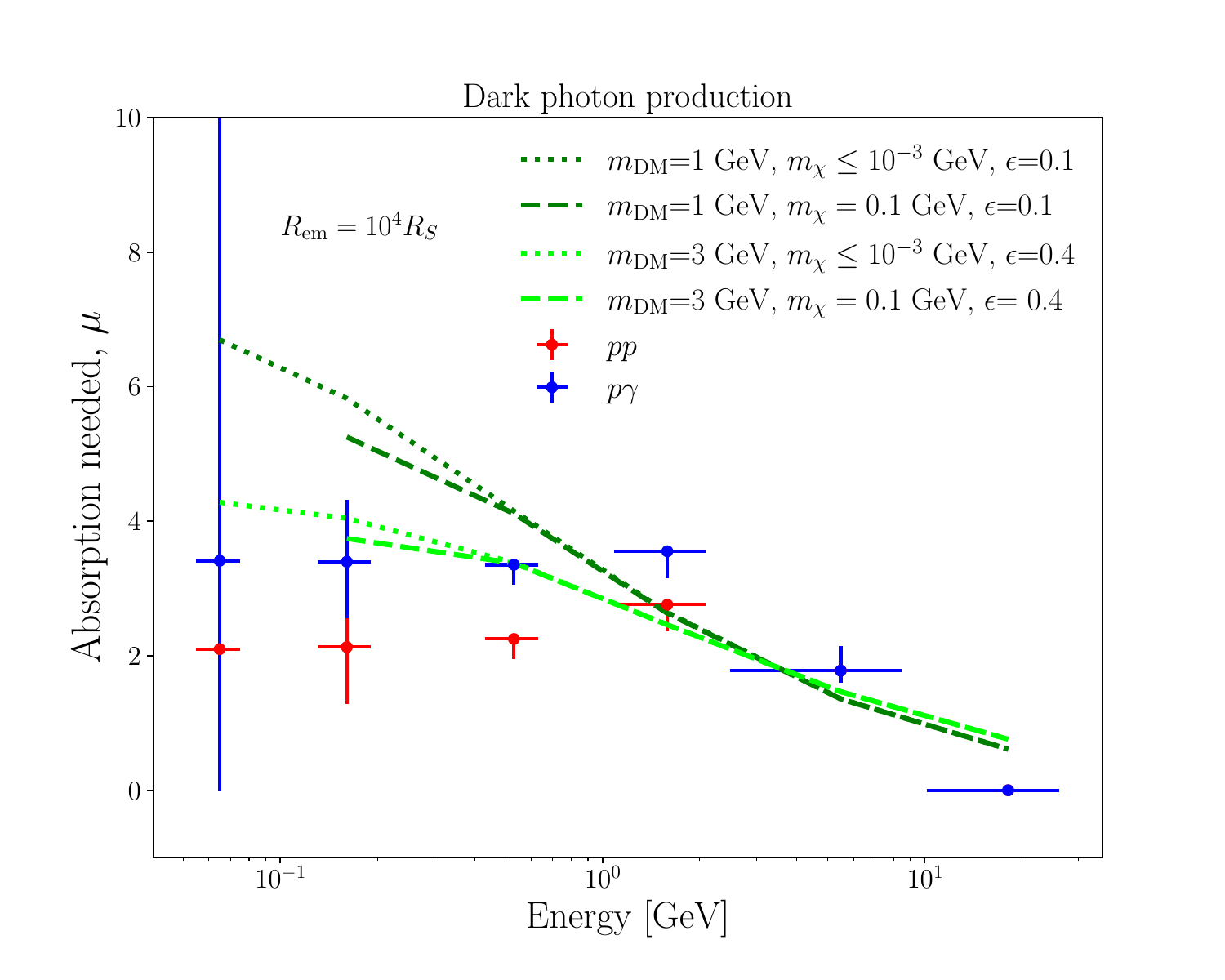}
		\includegraphics[width=0.45\textwidth]{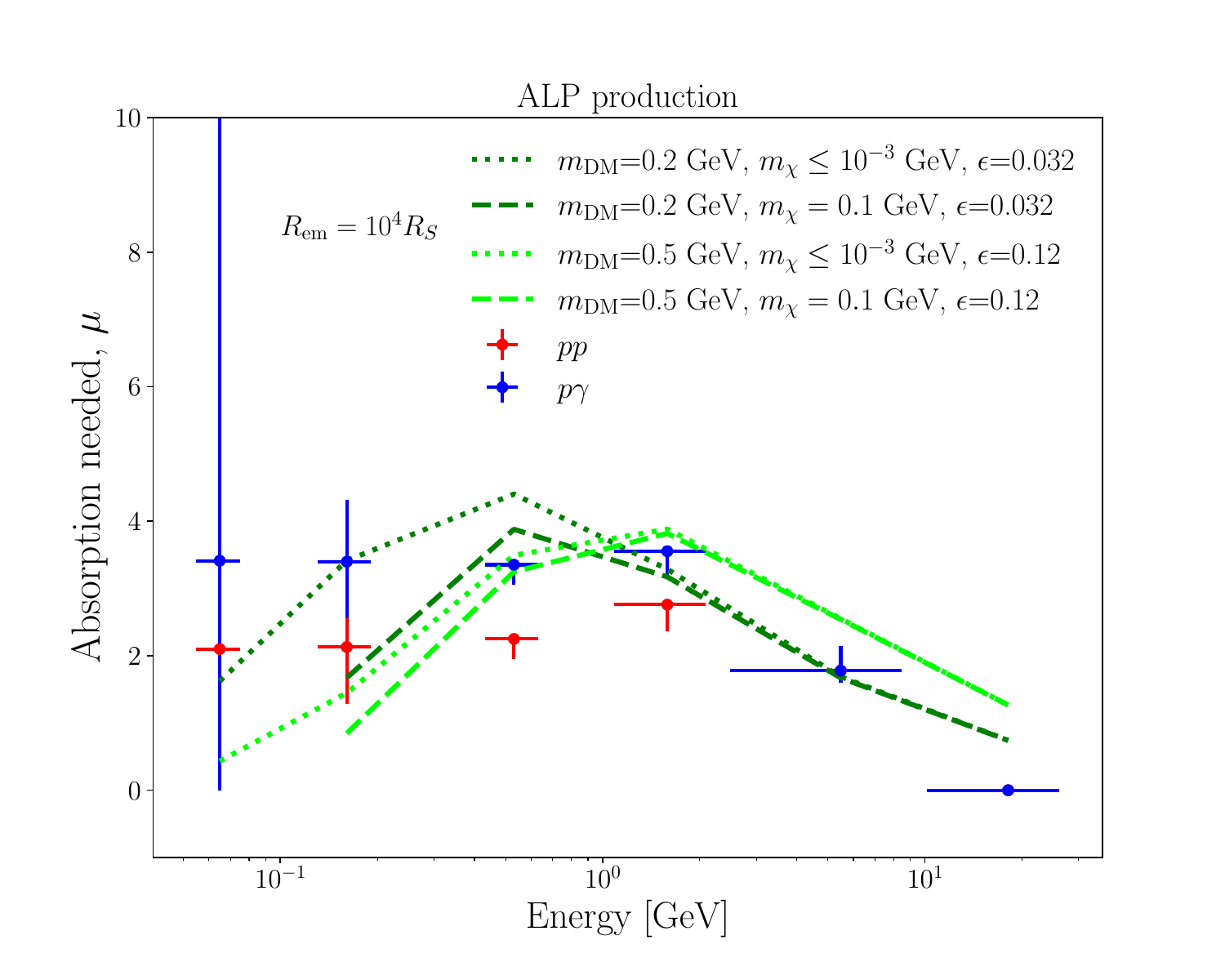}
		\centering
		\caption{Absorption coefficient needed to reconcile the photon flux produced by $pp$(red) and $p\gamma$(blue) interactions at $R_{\rm em}=10^{4}R_S$ with Fermi-LAT data, due to DM-photon inelastic scatterings inducing a dark photon (left panel) or an ALP (right panel) in the final state, as a function of the photon energy, for different values of the fermion dark matter mass, dark photon/ALP mass, and mixing.}
	\label{fig:attenuation-coeff_V_PS}
\end{figure*}
	
\section{Model-independent photon absorption by dark matter}
We begin by simply parametrizing the strength of the dark matter-photon interactions with a constant number $\sigma_{\rm DM-\gamma}$. Later in the manuscript we will refine this assumption and look into concrete scenarios. The column density of dark matter that photons travel through the spike of an AGN can be approximated by
\begin{align}
	\Sigma_{\rm DM}\big|_{\rm sp}& \simeq \int_{R_{\rm em}}^{R_{\rm sp}} dr  \rho_{\rm sp}(r)
\simeq \frac{\rho_0}{\gamma_{sp}-1} \frac{R_{sp}^{\gamma_{\rm sp}-\gamma}}{r_0^{-\gamma}R_{\rm em}^{\gamma_{sp}-1}}
	\label{eq:Sigma}
\end{align}
where we have used Eq.~(\ref{eq:spike_profile}) and that $R_{\rm em} \gg R_{\rm S}$. Expressed in terms of the different parameters
\begin{align}
    \Sigma_{\rm DM}& \simeq \left(\frac{4-\gamma}{5-\gamma}\right)M_{\rm BH}^{\frac{3-\gamma}{4-\gamma}}r_0^{\frac{\gamma}{4-\gamma}}R_{\rm em}^{-\frac{5-\gamma}{4-\gamma}}\alpha_{\gamma}^{-\frac{(-3+\gamma)^{2}}{-4+\gamma}}\rho_0^{\frac{1}{4-\gamma}}.
\end{align}
We can then analytically find the photon absorption coefficient induced by dark matter in NGC 1068. For instance, for an initial NFW profile with $\gamma=1$, we find

\begin{align}\label{eq:absorption_coeff}
\Delta\mu (E_\gamma)\simeq &~  
\left(\frac{M_{\rm BH}}{2\times 10^7 M_\odot}\right)^{2/3} \Big(\frac{r_0}{10\,{\rm kpc}}\Big)^{1/3}
\Big(\frac{\rho_0}{0.04 M_\odot/{\rm pc}^3}\Big)^{1/3} \nonumber \\ 
&\Big(\frac{R_{\rm em}}{10^3 R_{\rm S}}\Big)^{-4/3}\Big(\frac{m_{\rm DM}}{1\, {\rm GeV}}\Big)^{-1}
\left(\frac{\sigma_{{\rm DM}-\gamma}(E_\gamma)}{10^{-28}\,{\rm cm}^2}\right).
\end{align}

And, for $\gamma=2$
\begin{align}\label{eq:absorption_coeff_2}
\Delta\mu(E_\gamma)\simeq & ~ 
\left(\frac{M_{\rm BH}}{2\times 10^7 M_\odot}\right)^{3/4} 
\Big(\frac{r_0}{10\,{\rm kpc}}\Big)^{1/2}\Big(\frac{\rho_0}{0.4 M_\odot/{\rm pc}^3}\Big)^{3/8}\nonumber \\ 
&\Big(\frac{R_{\rm em}}{10^3 R_{\rm S}}\Big)^{-11/8}
\Big(\frac{m_{\rm DM}}{1\, {\rm GeV}}\Big)^{-1}
\left(\frac{\sigma_{{\rm DM}-\gamma}(E_\gamma)}{10^{-30}\,{\rm cm}^2}\right).
\end{align}

\begin{figure*}[t!]
		\centering
  		\includegraphics[width=0.45\textwidth]{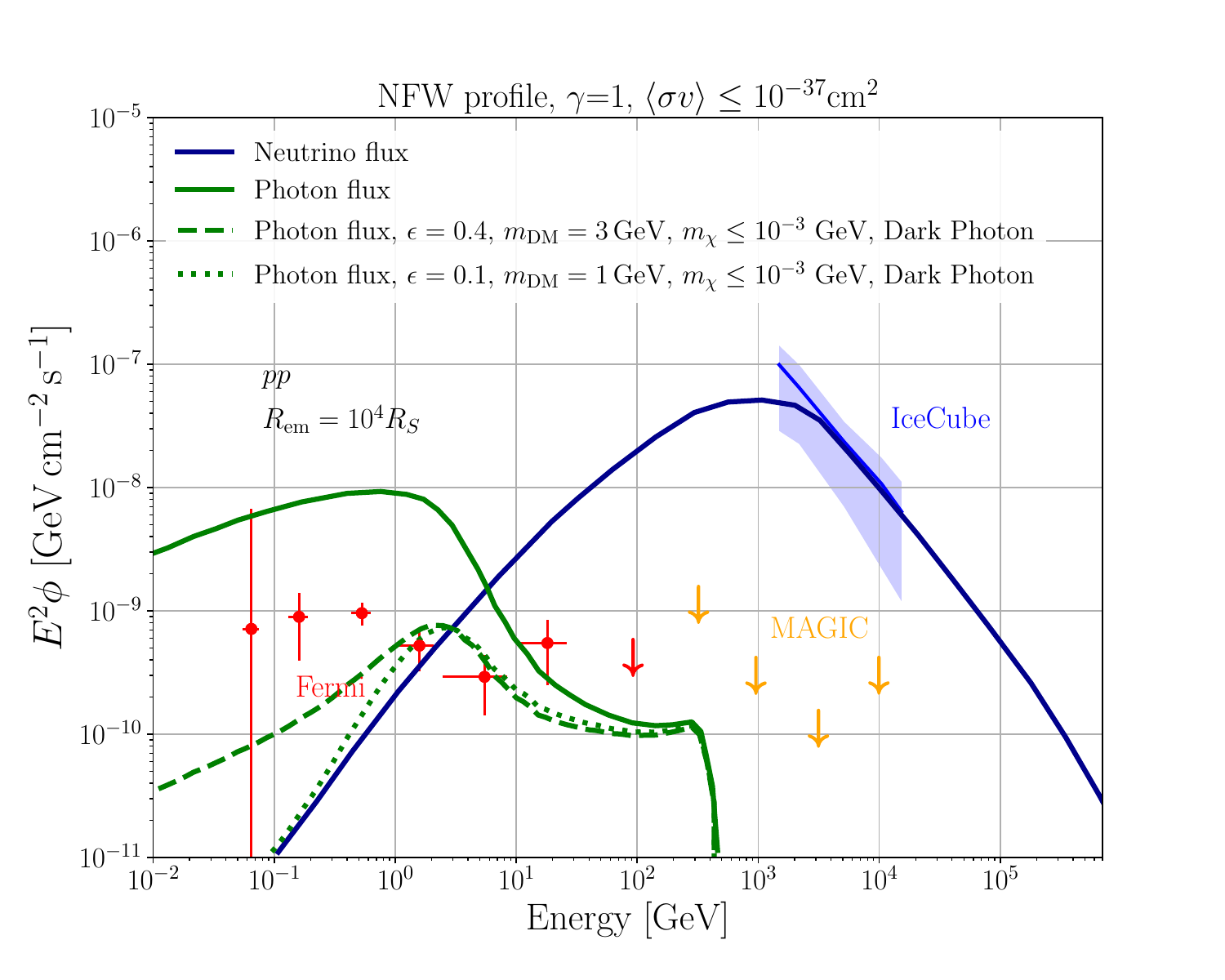}
  		\includegraphics[width=0.45\textwidth]{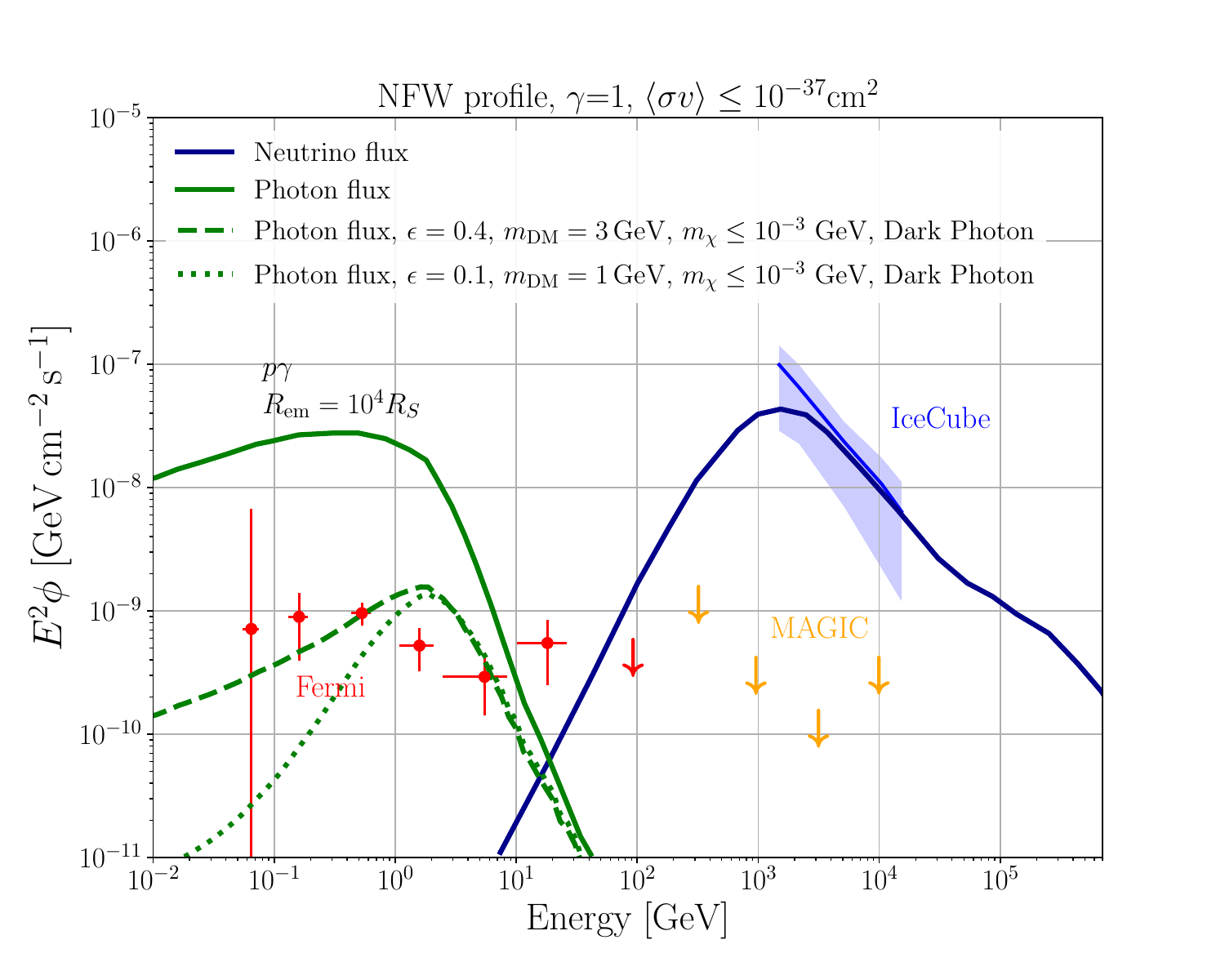}
        \includegraphics[width=0.45\textwidth]{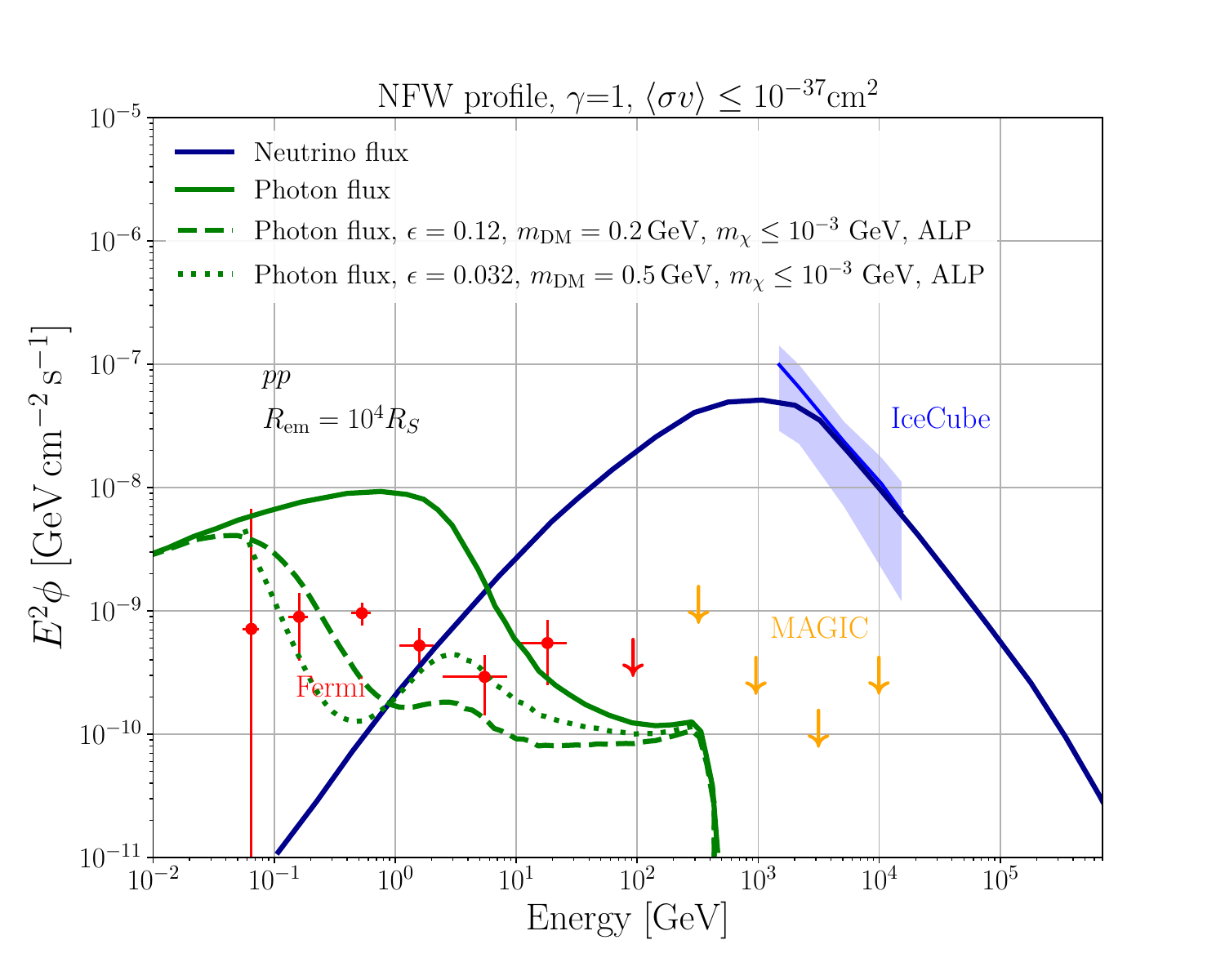}
		\includegraphics[width=0.45\textwidth]{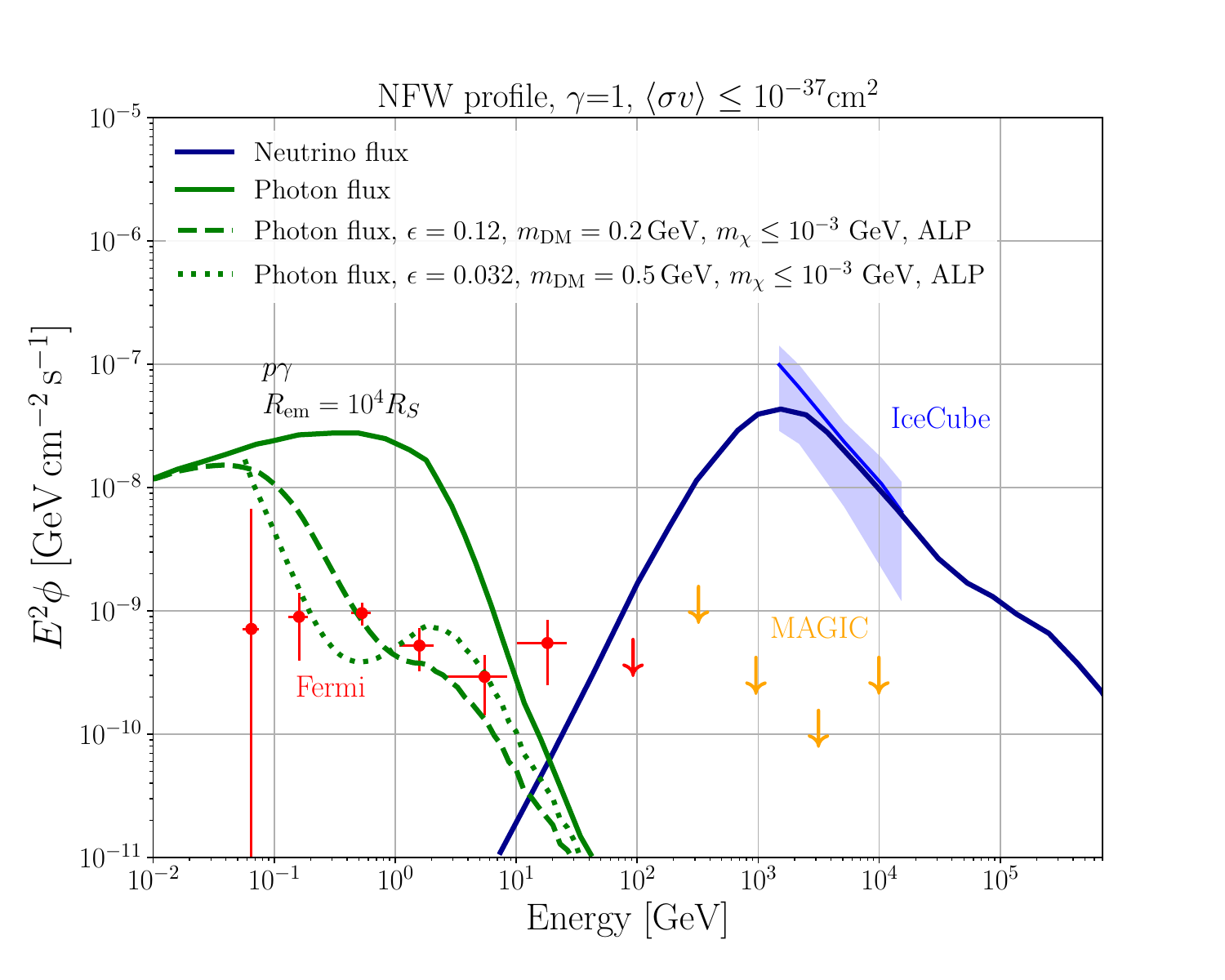}
		\centering
  		\caption{Spectral energy distribution from NGC 1068 for $pp$ processes (left panels) and $p\gamma$ processes (right panels) with emitting region $R_{\rm em} =10^{4}R_S$. We show the expected gamma-ray fluxes in astrophysical models and the corresponding attenuated fluxes induced by dark photon production (upper panels) and ALP production (lower panels), for different parameter choices inducing a sizable depletion of the gamma-ray flux.}
	\label{fig:attenuation-fluxes_PS_V}
    \end{figure*}

We thus conclude that for emission at $R_{\rm em}=10^{3}R_S$ there can be a significant absorption if the dark matter-photon interaction cross-section is at least $\sigma_{\rm DM-\gamma}/m_{\rm DM}\gtrsim 10^{-30}\,{\rm cm}^2/$GeV. Smaller values of the cross section may be possible for emitting regions closer to the supermassive black hole, or very cuspy dark matter spikes. Correspondingly, for farther emitting regions, larger values of the scattering cross section would be required.

We apply constant values of the dark matter-induced photon absorption cross section for high-energy neutrino and gamma-ray production via $p\gamma$ and $pp$ interactions in Figure \ref{fig:attenuation_constant}. For these models, as previously discussed, the high-energy neutrino flux is well fitted, but the predicted gamma-ray flux turns about one order of magnitude larger than Fermi-LAT data. When considering gamma-ray absorption induced by the surrounding dark matter with a cross section $\sigma_{\rm DM-\gamma}/m_{\rm DM} \simeq 10^{-28}$ cm$^2$/GeV, the ensuing fluxes are in clearly better agreement with Fermi-LAT data. As apparent from Eqs. \ref{eq:absorption_coeff} and \ref{eq:absorption_coeff_2}, the required cross section could be orders of magnitude smaller depending on the astrophysical parameters, crucially on the emitting region $R_{\rm em}$. As apparent from Eqs. \ref{eq:absorption_coeff} and \ref{eq:absorption_coeff_2}, one can see that the photon absorption coefficient scales approximately as 
\(\Delta\mu \propto R_{\mathrm{em}}^{-4/3}\)–\(R_{\mathrm{em}}^{-11/8}\), 
such that the required effective cross section satisfies 
\(\sigma_{\mathrm{DM}-\gamma}/m_{\mathrm{DM}} \propto R_{\mathrm{em}}^{4/3}\)–\(R_{\mathrm{em}}^{11/8}\). 
Hence, moving the emitting region from \(R_{\mathrm{em}}\simeq10^{4}R_{S}\) to \(R_{\mathrm{em}}\simeq10^{3}R_{S}\) (a range compatible with the expected coronal sizes of AGN) increases the local dark matter column density by roughly two orders of magnitude, allowing the same level of attenuation for a cross section smaller by a similar factor. 
Analogously, adopting a cuspier pre-existing halo profile (\(\gamma\gtrsim1\)) enhances the spike density and likewise reduces the required interaction strength. Hence, the quoted range \(\sigma_{\mathrm{DM}-\gamma}/m_{\mathrm{DM}}\sim10^{-28}-10^{-30}\,\mathrm{cm^{2}/GeV}\) should be interpreted as an effective value encompassing these astrophysical uncertainties rather than as an intrinsic theoretical ambiguity.

\section{Inelastic dark matter-photon scatterings}
The attenuation of the photon flux from NGC 1068 due to scatterings by dark matter particles will depend with the incoming photon energy in concrete models. For neutrino and gamma-ray production in the corona, from Figure \ref{fig:attenuation-coeff} one can see that for both $pp$ and $p\gamma$ models the absorption coefficient needed to reproduce Fermi-LAT data roughly decreases linearly with the photon energy. On the other hand, for production near $R_{\rm em} \sim 10^{4}R_S$, the absorption coefficient remains constant from $\sim 0.07-2 $ GeV, and decreases linearly to zero from $\sim 2-20$ GeV. In this section, we will focus on photon production at $R_{\rm em} \simeq 10^{4}R_S$, and we will discuss some scenarios where the dark matter particle can induce the required attenuation of the fluxes to explain Fermi-LAT data. In particular, we are interested in processes of the form
\begin{align}\label{eq:main_process}
\rm DM + \gamma  \rightarrow \rm DM + \chi
\end{align}
where $\gamma$ is the incoming photon flux, $\mathrm{DM}$ is the dark matter fermion present in the spike and halo of the galaxy, and $\chi$ is either a dark photon or an Axion Like Particle (ALP) coupled to the dark matter fermion. The cross section for the inelastic scattering process reads \cite{PhysRevD.18.3605, Brodsky:1986mi, Gondolo:2008dd, Dent:2019ueq}
\begin{align}
\sigma=\frac{\alpha g_{\chi}^2 \epsilon^{2}}{8 s} \frac{p}{k}\left[A(s)+B(s) \frac{\sqrt{s}}{p} \log \frac{2 p_0 k_0+2 p k-m_\chi^2}{2 p_0 k_0-2 p k-m_\chi^2}\right]
\end{align}
where $g_{\chi}$ is the dark boson $\chi$ coupling to the dark matter fermion $\mathrm{DM}$. We consider a fairly strongly coupled dark sector with $g_{\chi}=\pi$ \footnote{A strongly coupled dark sector prevents the final dark states $\chi$ to efficiently decay back into photons, which would spoil our purpose. The $\chi$ particles will rather undergo further scatterings with the dark matter fermion DM, losing energy on the process. This is due to the cross section for $\chi-\mathrm{DM}$ interactions scaling with $g_{\chi}^4$, while $\chi-e$ or $\chi-\gamma$ processes are suppressed by $\epsilon^2$.}. The  mixing connecting the dark sector to the visible sector via a dark photon mediator is denoted by $\epsilon$, and will be treated as a free parameter. $s$ denotes the center of mass energy of the scattering process. $A(s)$ and $B(s)$ depend on the nature of the final dark state and are provided in \cite{Gondolo:2008dd}, and $p_0=\left(s-m_{\rm DM}^2+m_\chi^2\right) / 2 \sqrt{s}$, $p=\left(p_0^2-m_\chi^2\right)^{1 / 2}, k_0=\left(s+m_{\rm DM}^2\right) / 2 \sqrt{s}$, and $k=\sqrt{s}-k_0$. In this work, we will focus on two concrete scenarios: a dark photon in the final state, and an axion like particle (ALP) in the final state.
\subsection{Dark photon production}
	\begin{figure*}[t!]
		\centering
		\includegraphics[width=0.45\textwidth]{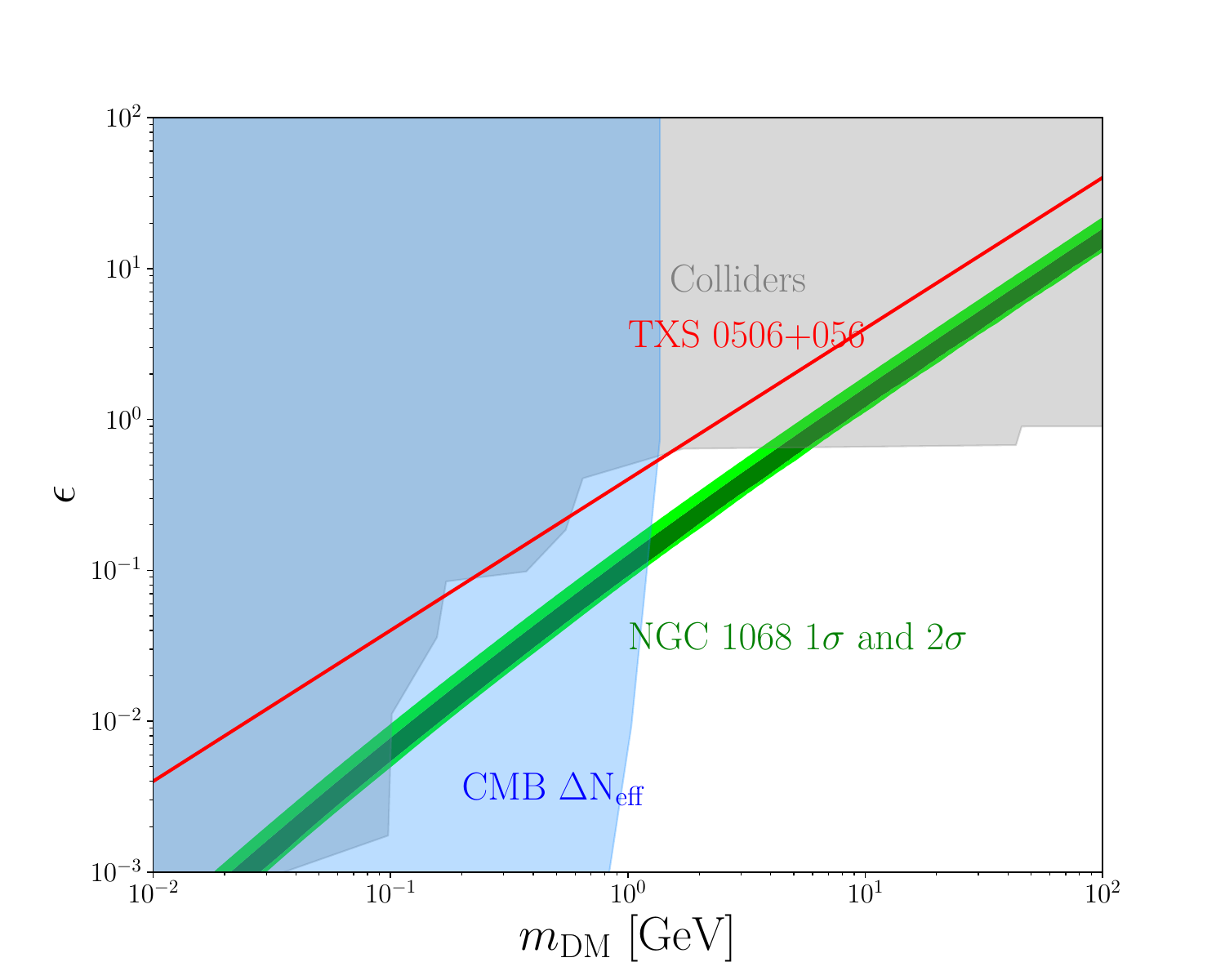}
		\includegraphics[width=0.45\textwidth]{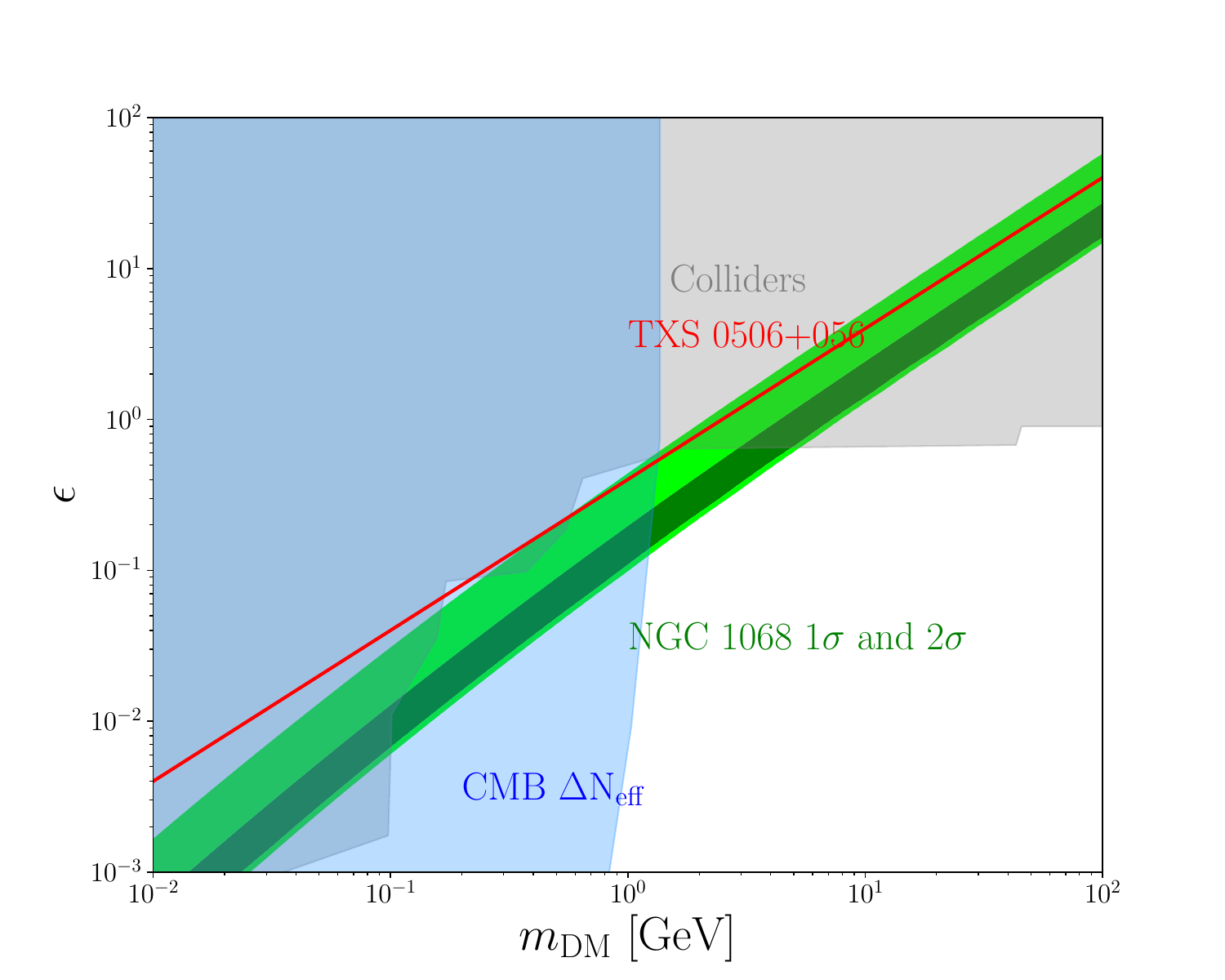}
		\centering
		\caption{Favored region at 1$\sigma$ (dark green) and 2$\sigma$ (green) of the parameter space of dark matter-photon interactions from Fermi-LAT and IceCube measurements of NGC 1068, for gamma-ray and neutrino emission via $pp$ processes (left panel) and $p\gamma$ processes (right panel). The parameter space is spanned by the dark matter mass $m_{\rm DM}$ and the kinetic mixing $\epsilon$ of the Standard Model photon to a new very light or massless dark photon mediator. For comparison, we show complementary constraints from colliders, cosmology and TXS 0506+056 (see main text for details).}
	\label{fig:favored_region}
\end{figure*}
In this case, we have
\begin{align}
A(s) = 2+\frac{2\left(m_{\rm DM}^2-m_\chi^2\right)}{s}+\frac{16\left(m_\chi^2+2 m_{\rm DM}^2\right) s}{\left(s-m_{\rm DM}^2\right)^2}
\end{align}
\begin{align}
B(s) = 2-\frac{4\left(m_\chi^2+2 m_{\rm DM}^2\right)}{s-m_{\rm DM}^2}-\frac{4\left(4 m_{\rm DM}^4-m_\chi^4\right)}{\left(s-m_{\rm DM}^2\right)^2}.
\end{align}

\begin{table*}[t]
\centering
\begin{tabular}{lcccccc}
\hline
Gamma-ray flux [$10^{-10}$ GeV cm$^{-2}$ s$^{-1}$] & 0.06 GeV & 0.15 GeV & 0.5 GeV & 1.5 GeV & 5 GeV & 15 GeV \\
\hline
Observed & $7.1^{+60}_{-7.0}$ 
         & $8.9^{+4.7}_{-5.0}$ 
         & $9.8^{+2.0}_{-2.4}$ 
         & $5.4^{+2.0}_{-2.1}$ 
         & $3.0^{+1.3}_{-1.6}$ 
         & $5.6^{+2.8}_{-3.0}$ \\
\hline
Expected $p\gamma$ & 220 & 270 & 270 & 170 & 16 & 0.71 \\
Attenuated $p\gamma$ (constant $\sigma_{\rm DM-\gamma}$) & 11 & 13 & 14 & 9.0 & 0.91 & 0.020 \\
Attenuated $p\gamma$ (dark photon) & 2.9 & 4.3 & 9.0 & 14 & 4.3 & 0.42 \\
Attenuated $p\gamma$ (ALP) & 130 & 13 & 3.8 & 6.9 & 2.9 & 0.40 \\
\hline
Expected $pp$ & 59 & 75 & 90 & 81 & 15 & 2.8 \\
Attenuated $pp$ (constant $\sigma_{\rm DM-\gamma}$) & 8.1 & 10 & 12 & 11 & 2.3 & 0.40 \\
Attenuated $pp$ (dark photon) & 0.79 & 1.4 & 3.3 & 7.0 & 4.2 & 1.3 \\
Attenuated $pp$ (ALP) & 41 & 4.5 & 1.4 & 3.4 & 3.2 & 1.4 \\
\hline
\end{tabular}
\caption{Comparison of observed and expected ($pp$ and $p\gamma$) gamma-ray fluxes at the relevant energies of Fermi-LAT, together with the attenuated fluxes induced by different scenarios: a constant dark matter-photon scattering cross section ($\sigma_{\mathrm{DM}-\gamma} / m_{\mathrm{DM}}=1.5 \times 10^{-28} \mathrm{~cm}^2 / \mathrm{GeV})$, the inelastic production of a dark photon with parameters $m_{\rm DM}=3$ GeV, $\epsilon=0.4$ and $m_{\chi} \ll m_{\rm DM}$, and the inelastic production of an ALP with parameters $m_{\rm DM}=0.5$ GeV, $\epsilon = 0.032$, and $m_{\chi} \ll m_{\rm DM}$. The data from this Table is illustrated in Figure \ref{fig:attenuation_constant} and Figure \ref{fig:attenuation-fluxes_PS_V}.}
\label{tab:data_fits}
\end{table*}

In the left panel of Figure \ref{fig:attenuation-coeff_V_PS}, we show the absorption coefficient obtained for the inelastic dark matter-photon scattering inducing a dark photon in the final state, for four benchmark values of the dark matter and  dark photon parameters. First, we show the absorption coefficient for a dark matter fermion with mass of $m_{\rm DM}=1$ GeV, a very light or massless dark photon mediator, $m_{\chi} \leq 10^{-3}$ GeV, and with  mixing $\epsilon=0.1$ (dotted line), and the same scenario for a heavier dark photon, $m_{\chi} = 0.1$ GeV (dashed line). In this case, the cross section resembles the one for compton scattering in the regime $E_\gamma \sim m_{\rm DM}$, where the cross section decreases roughly linearly with the photon energy. Here, the absorption coefficient obtained adequately suppresses the photon flux to the Fermi-LAT data from $\sim 2-20$ GeV, while it is slightly large in the lower energy bins. When the dark photon is massive, the cross section mildly flattens at low energies.

We also show the absorption coefficient for a dark matter mass of $m_{\rm DM}=3$ GeV, same choices for the value of the mediator mass as before, and larger values of the  mixing $\epsilon=0.4$. Here, in the lower energy bins we are in the regime $m_{\rm DM} \gg E_{\gamma}$, and the cross section is roughly constant. On the other hand, at the high-energy bins the cross section decreases with increasing photon energy. This scenario accommodates well the emitted fluxes from NGC 1068 to Fermi-LAT data in all energy bins. The benchmark values of the  mixing for a fairly massive dark photon of $m_{\chi}=0.1$ GeV are in some tension with collider searches and electroweak precision observables \cite{Fabbrichesi:2020wbt}, and theoretical considerations on the size of  mixing and perturbativity \cite{Cline:2024wja}, but are allowed for a very light or massless dark photon mediator \cite{Davidson_2000,Fabbrichesi:2020wbt,Liu:2019knx}.

In the upper panels of Figure \ref{fig:attenuation-fluxes_PS_V}, we show the attenuated photon flux from NGC 1068 in dashed and dotted green lines, for the same parameters discussed in Figure \ref{fig:attenuation-coeff_V_PS}.
Again, it can be noticed that the inelastic production of dark photons can deplete the gamma-ray flux to the observable levels for different choices of the parameters, however, a mass of $m_{\rm DM}=3$ GeV seems to accommodate the low-energy bins better than $m_{\rm DM}=1$ GeV, and visually, the agreement seems better for $p\gamma$ processes than for $pp$ processes.
\subsection{Axion production}
In this case, we have
\begin{align}
A(s) = -3+\frac{m_{\rm DM}^2-m_\chi^2}{s}+\frac{8 m_\chi^2 s}{\left(s-m_{\rm DM}^2\right)^2}
\end{align}
\begin{align}
B(s) = 1+\frac{2\left(m_\chi^2-4 m_{\rm DM}^2\right)}{s-m_{\rm DM}^2}+\frac{2\left(m_\chi^4-6 m_\chi^2 m_{\rm DM}^2+8 m_{\rm DM}^4\right)}{\left(s-m_{\rm DM}^2\right)^2}
\end{align}
In the right panel of Figure \ref{fig:attenuation-coeff_V_PS}, we show the absorption coefficient obtained for the inelastic dark matter-photon scattering inducing an ALP in the final state, for four benchmark values of the dark matter and ALP parameters. We show the absorption coefficient for a dark matter fermion with mass of $m_{\rm DM}=0.2$ GeV, a very light or massless dark photon mediator, $m_{\chi} \leq 10^{-3}$ GeV, and with  mixing $\epsilon=0.032$ (dotted line). We also show the same scenario for a heavier dark photon, $m_{\chi} = 0.2$ GeV (dashed line). Here, the cross section peaks at $E_{\gamma}=2 m_{\rm DM}$, roughly linearly decreasing at lower and larger energies. The absorption coefficient obtained in this scenario adequately suppresses the photon flux to the Fermi-LAT observation from $\sim 0.2-20$ GeV, while it is slightly smaller than required in the low-energy bins. When the dark photon is massive, the cross section decreases more sharply at low energies than in the nearly massless case.

We also show the absorption coefficient for a dark matter mass of $m_{\rm DM}=0.5$ GeV, same choices for the value of the mediator mass as before, and larger values of the  mixing $\epsilon=0.12$. Similar conclusions apply, being the absorption even stronger in the high-energy bins and weaker in the lower ones. These scenarios are only in mild tension with $\Delta N_{\rm eff}$ constraints from the CMB \cite{Vogel:2013raa}.

We also display in the lower panels of Figure \ref{fig:attenuation-fluxes_PS_V} the attenuated photon flux from NGC 1068 via ALP production in dashed and dotted green lines, for the same parameters discussed in Figure \ref{fig:attenuation-coeff_V_PS}.
It can be noticed that the inelastic production of ALPs can deplete the gamma-ray flux to the observable levels effectively in the high-energy bins, however, the depletion in the low-energy bins is small. In this case, there seems to be no obvious preference for $m_{\rm DM}=0.2$ GeV over $m_{\rm DM}=0.5$ GeV, nor a preference for $pp$ versus $p\gamma$ processes.

\section{Favored region of parameter space and complementary constraints}

We now turn into finding the combination of dark matter mass $m_{\rm DM}$ and mixing $\epsilon$ that reconciles the $pp$ and $p\gamma$ models for emission at $R_{\rm em} \simeq 10^{4}R_S$ with Fermi-LAT data. For this purpose, we perform a binned $\chi^2$ least-squares spectral fit of the $pp$ and $p\gamma$ attenuated fluxes to Fermi-LAT data, accounting for the uncertainty in the measured gamma-ray fluxes. The $\chi^2$-function in each bin $i$ is defined as
\begin{equation}
\chi^2_i\left(m_{\rm DM}, \epsilon\right)= \frac{\left(\Phi_{ i}\left(m_{\rm DM}, \epsilon\right)-\Phi_{i}^{\mathrm{obs}}\right)^2}{\left(\delta \Phi^{\mathrm{obs}}_i\right)^2},
\end{equation}
where $\Phi_i^{\rm obs}$ is the observed flux in each bin, $\delta \Phi^{\mathrm{obs}}_i$ is the uncertainty, for which we take the closest error bar end to the expected data for given dark matter parameters $\Phi_i$. In Table \ref{tab:data_fits}, we show the flux $\Phi_i^{\rm obs}$ data from Fermi-LAT at the relevant energy bins, including error bars, and the corresponding expected gamma-ray fluxes $\Phi_{i}$ for the different models under consideration, and some benchmark parameters. We find the 1$\sigma$ and 2$\sigma$ sensitivity contours on $m_{\rm DM}$ and $\epsilon$ by means of Wilks theorem as \cite{Wilks:1938dza}
\begin{equation}
\chi^2(m_{\rm DM}, \epsilon) - \chi^2_{\rm min} \leq 
\begin{cases}
2.3 & (1\sigma) \\
6.18 & (2\sigma)
\end{cases}
\end{equation}
where $\chi^2=\sum_{i} \chi^2_i$ and $\chi^2_{\rm min}$ is the minimum obtained when sampling over the free parameters $(m_{\rm DM}, \epsilon)$. In reality, the $\chi^2$-function also depends on the dark gauge coupling $g_{\chi}$, and on several theoretical astrophysical parameters. As previously indicated, we fix the dark gauge coupling to $g_{\chi}=\pi$, and the astrophysical parameters to those of the concrete $pp$ and $p\gamma$ scenarios for $R_{\rm em}=10^{4}R_S$ discussed in section \ref{sec:fluxes}. We further focus here in the dark photon production scenario, which fits the spectral dependence of the observed gamma-ray deficit somewhat better than the ALP case. Our results are shown in Figure \ref{fig:favored_region} as green colored bands. The left panel shows the results for $pp$ processes, and the right panel for $p\gamma$. For comparison, we show bounds from colliders (in shaded grey color) \cite{Davidson_2000, Fabbrichesi:2020wbt} and CMB measurements of $\Delta N_{\rm eff}$ \cite{Vogel:2013raa, Adshead:2022ovo} (in shaded blue color). The preferred region of parameter space from NGC 1068 is allowed by colliders in the range of dark matter masses $m_{\rm DM} \simeq 0.1-5$ GeV, while measurements of $\Delta N_{\rm eff}$ disfavor a millicharged dark matter mass smaller than $m_{\rm DM} \lesssim 1$ GeV. Avenues to avoid this bound in non-standard cosmological histories are possible.

Additionally, we show constraints arising from the non-observation of a significant attenuation of gamma rays in TXS 0506+056 \cite{Ferrer:2022kei}. This bound is compatible with the preferred region in NGC 1068 within $2 \sigma$. Interestingly, TXS 0506+056 also suffers from overshooting X-ray data if aiming to fit the high-energy neutrino observations with single-zone lepto-hadronic models \cite{Keivani_2018, Gao:2018mnu}, so a similar mechanism to the one proposed for NGC 1068 may be at play. Slightly stronger bounds than these have been derived for Tidal Disruption Events in \cite{Fujiwara:2023lsv}. However, the high-energy neutrino association of these events is yet to be confirmed by IceCube, so such bounds are taken with a grain of salt in here.

\section{Outlook and Conclusions}

We have focused our discussion on dark matter masses in the range $m_{\rm DM} \simeq 0.01-10$ GeV, however, lighter dark matter particles would require smaller values of the kinetic mixing for a given photon absorption coefficient due to the enhancement in number density at the source, and may be allowed by cosmological bounds \cite{Vogel:2013raa}, and astrophysical bounds from \textit{e.g} stellar/supernova cooling \cite{Raffelt:1996wa}.
Furthermore, it is worth noting that those regions of parameter space in tension with complementary constraints might be allowed, for example, if more than one single scattering channel were open, or multiple final states were present. Furthermore, the dark matter column density could also be one to two orders of magnitude larger than customarily assumed in our analysis, \textit{e.g} if the pre-existing dark matter profile were cuspier than $\gamma \gtrsim 1$, or the emitting region of neutrinos and gamma rays lies in between the corona and the BLR, e.g $R_{\rm em} \sim 10R_S-10^{3}R_S$. For coronal emission, the absorption coefficients needed to reconcile the emitted photon flux with Fermi-LAT data are smaller than at $R_{\rm em}\sim 10^{4}R_S$, and the column density of dark matter particles is much larger, requiring lower couplings of the dark matter to the visible sector. This possibility will be thoroughly studied in a future analysis.

Although we have provided some concrete particle physics examples with the correct energy dependence of the cross section, we emphasize that a simple constant dark matter-photon cross section of $\sigma_{\rm DM-\gamma}/m_{\rm DM} \sim 10^{-28}-10^{-30} \rm \, cm^2/\mathrm{GeV}$ suppresses the expected fluxes adequately. Such values are in slight conflict with cosmological constraints \cite{B_hm_2014, Escudero_2018, Crumrine:2024sdn}, although these apply at much lower photon energies $ E_{\gamma }\lesssim 1$ keV, where the scattering cross section may be smaller than at the $ E_{\gamma} \sim$ GeV photon energies considered in this work. Further model-building efforts along these lines would be beneficial in confronting cosmological bounds with the preferred astrophysical cross section adequately.

There are related alternatives to the underlying idea of this work. We have focused on the attenuation of the gamma-ray flux due to dark matter-photon scatterings in the AGN. Since gamma rays are produced by interactions of accelerated protons and electrons at the source, a sizable coupling between the dark matter and protons/electrons would also cause indirectly a modification of the emitted gamma-ray fluxes. This possibility, already considered in the AGN context in \cite{Herrera:2023nww, Gustafson:2024aom, Mishra:2025juk, DeMarchi:2024riu, Wang:2025ztb}, will be studied as an alternative mechanism to deplete the gamma-ray flux from NGC 1068 to the observed level in future work.

Our work also has caveats. Importantly, there is a strong degeneracy in our setup between the column density of dark matter in the AGN and the dark matter-photon scattering cross section. To reduce such degeneracy, independent methods to infer the dark matter distribution in the vicinity of AGN would be needed, together with dedicated simulations of the dark matter distribution in the region of gravitational influence of supermassive black holes. Furthermore, translating a cross section of $\sigma_{\rm DM-\gamma}/m_{\rm DM} \sim 10^{-28}-10^{-30} \rm \, cm^2/\mathrm{GeV}$ into concrete models is not a trivial task. In our set-up, for instance, we require a fairly strongly coupled dark sector ($g_{\chi}\sim0.5-4\pi$)  in order to avoid collider and cosmological constraints in some regions of parameter space. However, such models may induce large self-interactions in galaxies and clusters of galaxies, a possibility tightly constrained \cite{Randall:2008ppe, Markevitch:2003at}. This, however, depends on the hierarchy of dark matter and dark photon/ALP masses. Another possibility would be to introduce a mass splitting between the initial and final dark matter states in Eq. \ref{eq:main_process}, which may prevent self-interactions for non-relativistic dark matter.

Our scenario is testable on multiple fronts. Future high-energy neutrino and gamma-ray data from AGN will help us better understand whether there is an emergent pattern on the ratio of gamma-ray to neutrino fluxes, and whether astrophysical models can accommodate the fluxes for motivated values of the astrophysical parameters. Several other AGN have been identified with IceCube at milder significance than NGC 1068 \cite{IceCube:2024dou}, and a dedicated comparative analysis of the gamma-ray and high-energy neutrino data from these sources will be pivotal to test lepto-hadronic models further. The preferred dark matter parameters inferred here shall apply universally, so multiple source observations should allow us to narrow down the allowed parameter space. Future collider and refined comological probes will also help to independently test the scenario presented here.

The observed gamma-ray obscureness of NGC 1068 may be due to an efficient acceleration mechanism of cosmic rays only within $\sim 30$ times the Schwarzschild radius of the central black hole, and the presence of a starburst region farther away with high supernova rates $\sim 0.5 \, \mathrm{yr}^{-1}$. This astrophysical scenario is a plausible simultaneous explanation of Fermi-LAT and IceCube data.
Here we have proposed yet another plausible scenario which we argue that reduces the amount of astrophysical free parameters and their level of fine-tuning. The price to pay is requiring the dark matter of the Universe to scatter with photons with a cross section of $\sigma_{\rm DM-\gamma}/m_{\rm DM} \simeq 10^{-28}-10^{-30}$ cm$^2$/GeV.
\begin{acknowledgments}
We are grateful to Kohta Murase for useful feedback on the high-energy neutrino and gamma-ray emission from NGC 1068. We also thank Alejandro Ibarra and Elisa Resconi for useful input at the early stages of this work. This work is supported by the U.S. Department of Energy under award numbers
DE-SC0020250 and DE-SC002026. We also thank the Technical University of Munich and the Max-Planck Institute for Particle Physics for funding and support in the early stages of this work.
\end{acknowledgments}

\bibliography{References}
\clearpage

\onecolumngrid
\appendix
\end{document}